# FPGA-Based Implicit-Explicit Real-time Simulation Solver for Railway Wireless Power Transfer with Nonlinear Magnetic Coupling Components

Han Xu, *Student Member, IEEE*, Yangbin Zeng, *Member, IEEE*, Jialin Zheng, *Graduate Member, IEEE*, Kainan Chen, *Member, IEEE*, Weicheng Liu, *Student Member, IEEE*, Zhengming Zhao, *Fellow, IEEE*

*Abstract*—Railway Wireless Power Transfer (WPT) is a promising non-contact power supply solution, but constructing prototypes for controller testing can be both costly and unsafe. Real-time hardware-in-the-loop simulation is an effective and secure testing tool, but simulating the dynamic charging process of railway WPT systems is challenging due to the continuous changes in the nonlinear magnetic coupling components. To address this challenge, we propose an FPGA-based half-step implicit-explicit (IMEX) simulation solver. The proposed solver adopts an IMEX algorithm to solve the piecewise linear and nonlinear parts of the system separately, which enables FPGAs to solve nonlinear components while achieving high numerical stability. Additionally, we divide a complete integration step into two half-steps to reduce computational time delays. Our proposed method offers a promising solution for the real-time simulation of railway WPT systems. The novelty of our approach lies in the use of the IMEX algorithm and the half-step integration method, which significantly improves the accuracy and efficiency of the simulation. Our simulations and experiments demonstrate the effectiveness and accuracy of the proposed solver, which provides a new approach for simulating and optimizing railway WPT systems with nonlinear magnetic coupling components.

## I. INTRODUCTION

RAILWAY wireless power transfer (WPT) systems emerge as a promising non-contact power supply solution, effectively tackling the reliability concerns linked to contact-based power transfer [1]. Investigations into various control issues are essential in the development of wireless power transfer systems. However, constructing prototypes for controller testing incurs high costs and safety concerns. Recently, real-time hardware-in-the-loop (RT-HIL) simulation serves as an affordable and secure testing tool to provide virtual testing environment for controllers [2].

Accurately simulating the targeted system under different operation conditions is crucial for RT-HIL simulation. In the case of WPT systems, a significant operational condition is the dynamic transitions of the receiver coils between different transmitter coils causing nonlinear changes in the magnetic coupling component while the train is in motion [3]. Additionally, the high switching frequency of Railway WPT systems requires a small simulation time step to simulate the dynamic system behaviors. To achieve small step-sizes, field-programmable gate array (FPGA)-based simulators are preferred for their low computational latency [4]. However, it is still difficult to solve the nonlinear (NL) variations within such limited time steps on FPGAs.

To address the aforementioned issues, three approaches exist for solving the NL power electronics system (PES): explicit methods, implicit iterative methods, and latency-based decoupling methods. In [5], [6], explicit methods are utilized to solve the PES that contains NL components. In the explicit methods, the integration only depends on the solutions at previous time steps, so there is no need to solve NL equations and small step-sizes can be achieved. However, these methods may lead to unstable results, due to the limited numerical stability.

Compared with explicit methods, implicit iterative methods have better numerical stability [7], but require iteratively solving NL equations per time step, causing large computation delays and nonuniform computation times, which hinder achieving small step-sizes [8]. In [9], [10], compensation methods are further adopted to reduce computational burdens of the iterative process, but small step-sizes are still difficult to be realized.

The main concept of latency-based decoupling methods is to partition the whole system into NL and piecewise linear (PWL) parts and then solve them with different methods or step-sizes to avoid solving NL equations. In [11], [12], the NL parts are assumed to be much slower than the PWL parts. Under this assumption, interface variables of the NL parts are treated as constant sources for multi-steps. However, when the changing rates of NL parts are fast, this method will introduce large numerical errors. Additionally, the latency-based linear multi-step compound method (LB-LMC) is proposed in [13], and used in FPGA-based RT-HIL simulations [14]–[16]. In this method, NL and PWL parts are solved with explicit and implicit methods respectively and the same step-size. Therefore, the assumption that the NL parts are slow is not needed, and solving NL equations is avoided. However, this method introduces a single-step delay between the PWL and NL parts. This inherent one-step delay significantly compromises the numerical stability [17]. The compromised numerical stability necessitates the adoption of exceedingly small-time steps during the solution of certain systems to avoid divergence, leading to substantial limitations on the range of solvable systems and their associated parameters. In the specific context of solving the WPT system targeted in this

This work was supported by the National Natural Science Foundation of China under Grants 52207209 and U2034201, and funded by China Postdoctoral Science Foundation under Grant 2021M701844. (Corresponding author: Yangbin Zeng.)

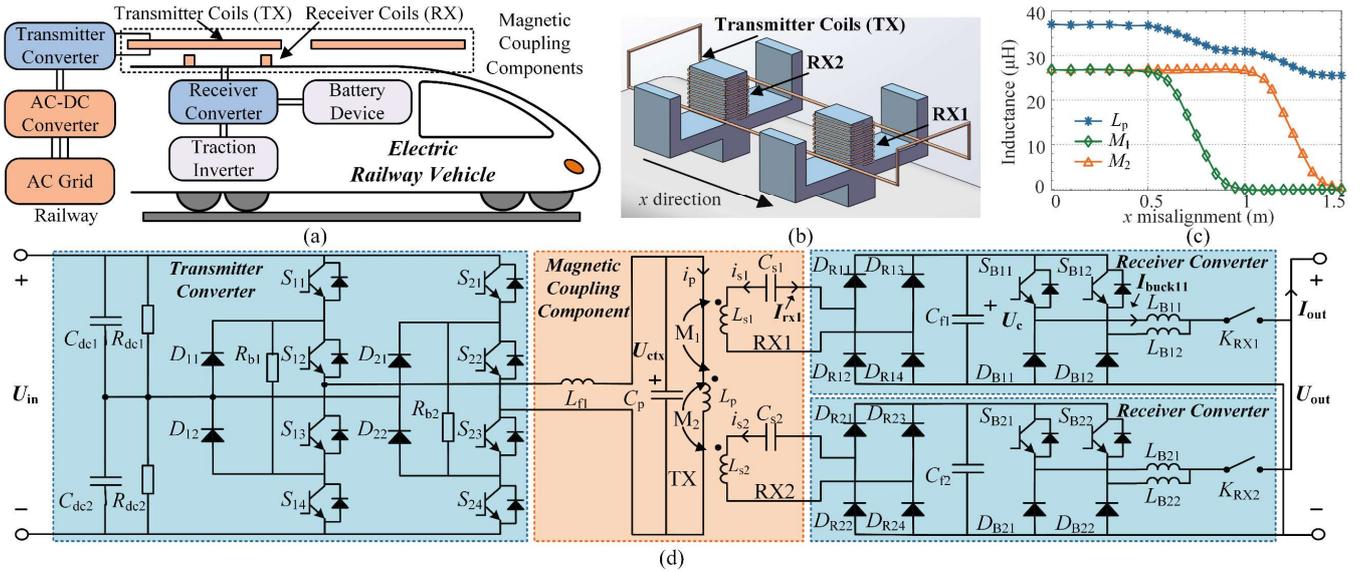

Fig. 1. Railway WPT system. (a) Diagram of railway WPT System. (b) Magnetic coupling component model. (c) Inductance values v.s. *x* misalignment. (d) WPT system circuit diagram.

paper, the application of this approach results in divergence, but our proposed method converges under the same step-size.

In summary, explicit methods can avoid solving NL equations to achieve small step-sizes, while implicit methods have good stability but involve iterative NL equation solving, causing large computational delays. Latency-based decoupling methods separate NL and PWL parts, aiming to avoid solving NL equations and achieve numerical stability. However, the one-step delay still causes instability issues and low accuracy order. In fact, the studied WPT system in this paper contains resonant capacitors with small values, making the latency-based method diverges due to the limited stability.

To address this, an FPGA-based implicit-explicit (IMEX) simulation solver is proposed for solving railway WPT with NL magnetic coupling components. After dividing the system into NL and PWL parts, the NL and PWL parts are solved respectively with explicit and implicit methods. Different from latency-based methods, this method eliminates the one-step delay by dividing one integration step into two computational stages. The latter stage eliminates the one-step delay, and increases the numerical accuracy and stability compared to conventional latency-based methods. As a result, this solver can avoid solving NL equations, enabling FPGAs to solve NL components, and also achieve high numerical accuracy and stability.

The primary contributions of this paper are as follows:
1. An efficient FPGA-based IMEX solver is proposed to solve WPT systems containing nonlinear magnetic coupling component.
2. The solver is implemented on the FPGA to solve a WPT system with a step-size of 75ns.
3. The solver provides a potential solution to simulating other power electronics systems containing nonlinear components.

The remain of this paper is organized as follows. Section II introduces the studied Railway WPT system including its modelling and characteristics. Section III delves into the proposed IMEX simulation solver. Section IV introduces the FPGA implementation results. Section IV presents the case study on the railway WPT system and evaluates the advantages of the proposed solver. Section V concludes this paper.

## II. RAILWAY WPT SYSTEM
### A. System Overview

This section introduces a simulated 350-kW railway WPT system, comprising a transmitter unit, magnetic coupling unit, and receiver unit, as shown in Fig. 1(a). The transmitter unit derives electrical energy from the traction power system and employs the AC-DC-AC converter to supply high-frequency AC power to the magnetic coupling unit. And then the magnetic coupling unit, facilitated by a coil loose coupling device, achieves wireless power transmission. The receiver unit rectifies the high-frequency AC power onto the DC bus, subsequently distributing it separately to the traction inverter and the onboard battery unit.

The battery can be charged wirelessly through the WPT system during the train in motion. The magnetic coupling unit includes transmitter and receiver coils, with the transmitter coil fixed above the train and the receiver coil fixed on the top of the train, as shown in Fig. 1(b). A high-speed train consists of multiple receiving coils, while the transmitting coil is also divided into multiple segments. As a result, during the dynamic operation of the train, the receiving coils continuously move in and out of the transmitting coil's vicinity. This leads to variations in the self-inductance and mutual inductance of the receiving and transmitting coils that change as the train travels.

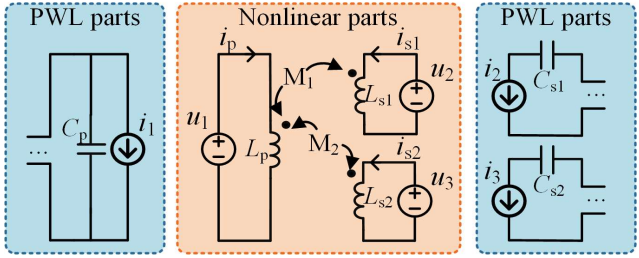

Fig. 2. Railway WPT system partitioning.

To obtain accurate changes in self-inductance and mutual inductance with respective to positions, finite element simulation tools are adopted. The variations in mutual inductance and self-inductance as the train's receiver coil progressively moves away from a particular section of the transmitter coil are shown in Fig. 1(c). The variations are obtained through finite element simulation. By altering the position of the receiving coil relative to the transmitting coil, we simulated a total of 26 data points to reveal the relationship between inductance and position.

As depicted in Fig. 1(c), it can be observed that as the receiving coil moves along the x-direction indicated in Fig. 1(b), the receiving coil initially moves away from the envelope region of the transmitting coil, leading to a decrease in the mutual inductance. Concurrently, the self-inductance of the transmitting coil also decreases. Continuing in the $x$-direction, the receiving coil similarly moves farther away from the transmitting coil's envelope region, causing the mutual inductance $M_2$ to begin decreasing. Simultaneously, the self-inductance $L_p$ of the transmitting coil continues to decrease. Similarly, when the receiving coil enters the envelope region of the next transmitting coil, the mutual inductance starts to increase. In summary, within the context of the railway WPT system, the continuous positional changes between the transmitting and receiving coils during train motion result in dynamic changes in the mutual inductance within the magnetic coupling component.

As the efficiency of power transfer in the WPT system's magnetic coupling unit is significantly influenced by the mutual inductance between the transmitting and receiving coils, the wireless power transmission of the WPT system experiences notable fluctuations during the train's movement. Consequently, it becomes essential to investigate corresponding control strategies to address these variations, thereby achieving high-speed, efficient, and reliable power transmission in the railway WPT system.

Fig. 1(d) shows the circuit diagram of the WPT system. In the context of the railway WPT system, the transmitter converter employs a passive clamp three-level converter. The magnetic coupling unit utilizes a parallel-series compensating topology, where a single transmitting coil can accommodate two receiving coils along with their corresponding converters. The receiver unit employs uncontrolled rectification, leading to the use of two parallel-connected buck converters for voltage regulation in each receiver unit. From a circuit perspective, the coupling inductance and self-inductance of the magnetic coupling component in the WPT circuit change as the train moves. The nonlinear dynamic variations in the self-inductance and mutual inductance while the train in motion influences the performance of the WPT system. Therefore, testing the systems solely in static environments might overlook the unexpected issues in real environments.

*B. Mathematical Characteristics of the System*

Due to the challenges of testing in a static environment, HIL simulation can become a good alternative for testing WPT systems. However, supporting WPT systems with HIL simulation is difficult due to the following three reasons: from the perspectives of algorithms, components, and step size.

1) **Stiffness of PWL part**: As shown in Fig. 1(d), the simulated system's receiver circuit includes resonant capacitors $C_{s1}$, $C_{s2}$ with a small value (62.34nF) that introduces fast-changing variables. From a mathematical perspective, the system's corresponding state equation has eigenvalues with large magnitudes. If an algorithm with a small stability region is used to solve the system, the simulation results may diverge.

2) **Nonlinearity of NL part**: As depicted in Fig. 1(c), the inductance of the magnetic coupling component undergoes continuous NL changes when the train is in motion. As the inductance values continuously vary, it becomes impractical to store all the corresponding equations. However, solving the NL equations iteratively makes it challenging to achieve small step-sizes due to the large computational delays.

3) **High Switching Frequency**: The maximum switching frequency in the system is 40kHz. To provide accurate simulation results, the step-size needs to be as small as possible. Generally, a step size of 250ns can provide relatively accurate simulation results at this switching frequency [18].

III. MODELLING AND SOLVER DESIGN

*A. Modeling of the Railway WPT System*

Based on the above analysis, NL and PWL parts have different mathematical characteristics and bring out different challenges for simulation. Therefore, our method partitions the whole WPT system into different subsystems so that their characteristics can be tackled with different methods. As illustrated in Fig. 2, the magnetic coupling component belongs to the NL part, while the rest of the power electronics circuits lie in the PWL part.

This paper employs the state equation method for solving the system. For generating the complete state equations of the WPT system, the NL and PWL parts can be considered as being connected through pairs of equivalent sources. The current of the corresponding voltage sources determines the controlled current sources, while the voltage of the corresponding current sources determines the equivalent controlled voltage sources. This model is completely equivalent to the original circuit because pairs of controlled sources are equivalent to ideal transformer models (ITM) with turns ratio being one.

The state and output equations of the NL parts can be represented in a general form as
$$\dot{x}_{nl} = f_{nl}(x_{nl}, y_1), \quad (1)$$

$$y_{nl} = g_{nl}(x_{nl}) = x_{nl}. \quad (2)$$

The NL parts of the WPT system can be represented by substituting the general state function $f_{nl}$ and output function $g_{nl}$ with the following functions:

$$f_{nl}(x_{nl}, y_l) = y_l, \quad g_{nl}(x_{nl}) = M^{-1}x_{nl}$$

$$M^{-1} = \frac{1}{L_{s2}M_1^2 + L_{s1}M_2^2 - L_p L_{s1} L_{s2}}, \quad (3)$$

$$\begin{pmatrix} -L_{s1}L_{s2} & L_{s2}M_1 & L_{s1}M_2 \\ L_{s2}M_1 & M_2^2 - L_p L_{s2} & -M_1 M_2 \\ L_{s1}M_2 & -M_1 M_2 & M_1^2 - L_p L_{s1} \end{pmatrix}$$

where $x_{nl}=[\Psi_p; \Psi_{s1}; \Psi_{s2}]$, $\Psi_p$, $\Psi_{s1}$, $\Psi_{s2}$ are the magnetic flux of inductors $L_p$, $L_{s1}$ and $L_{s2}$, $y_{nl}=[i_1;i_2;i_3]$, $y_l=[u_1;u_2;u_3]$, and these variables are defined in Fig. 2. Inductors $L_p$, $L_{s1}$, $L_{s2}$, $M_1$, $M_2$ are nonlinear inductors that are modeled as functions of relative position $x$ and inductor currents.

The state and output equation of the PWL parts can be written as

$$\dot{x}_l = A_k x_l + \begin{bmatrix} B_k^1 & \vdots & B_k^2 \end{bmatrix} \begin{bmatrix} u_l \\ \cdots \\ y_{nl} \end{bmatrix}, \quad (4)$$

$$y_l = C_k x_l + \begin{bmatrix} D_k^1 & \vdots & D_k^2 \end{bmatrix} \begin{bmatrix} u_l \\ \cdots \\ y_{nl} \end{bmatrix}, \quad (5)$$

where the subscript l identifies the PWL parts, $u_l$ is a vector of independent sources (e.g., $U_{in}$ in this specific case), $x_l$ is a vector of independent state variables (e.g., the independent capacitor voltage and inductor current), $A_k, B_k^1, B_k^2, C_k, D_k^1, D_k^2$ are system matrices that correspond to the $k$-th switching state. The detailed process for constructing these system matrices can be found in chapter 8 of [19]. Moreover, the semi–symbolic state equation generation method [20] is adopted to facilitate constructing system matrices. Using this way, a unified way of representing system matrices corresponding to different topologies can be achieved.

*B. FPGA-based IMEX Simulation Solver*

After the state equations are partitioned into NL and PWL parts, an IMEX simulation solver is proposed to solve these equations. This solver utilizes the aforementioned partitioning results to solve NL and PWL parts with different methods, all the while avoiding introducing one-step delay or compromising accuracy. In this solver, each complete integration step consists of two different stages and the NL and PWL parts are solved in parallel, as shown in Fig. 3. Assuming a time step of $h$ and the system is at current point $t_n$, the computational process of this solver is outlined below.

Stage 1: The system integrates from current time point $t_n$ to midpoint $t_{n+1/2}=t_n+h/2$.

Stage 1-1: NL and PWL parts parallelly calculate the interface variables $y_l$ and $y_{nl}$ at time point $t_n$ through (2) and (5).

Stage 1-2: Upon obtaining the interface variables, the NL and PWL parts are stepped in parallel from $t_n$ to $t_{n+1/2}$ through

$$x_{nl}(t_{n+1/2}) = x_{nl}(t_n) + h/2 f_{nl}(x_{nl}(t_n), y_l(t_n)), \quad (6)$$

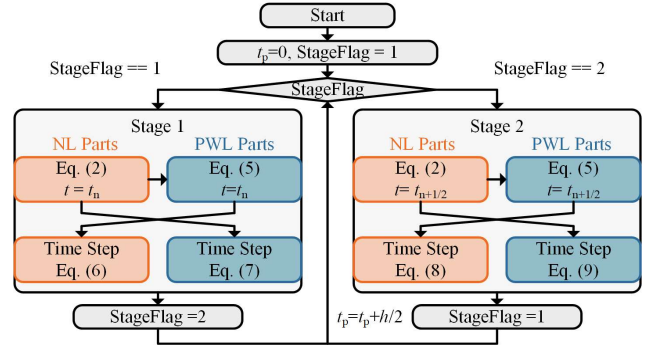

Fig. 3. Flowchart of the proposed solver.

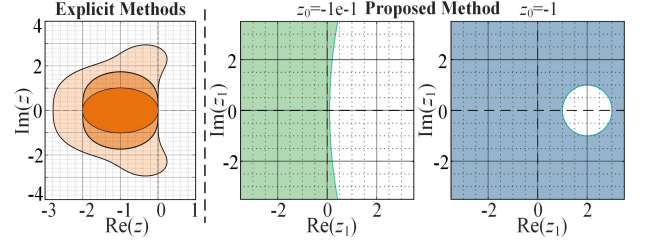

Fig. 4. Stability regions with different $z_0$.

$$x_l(t_{n+1/2}) = (I - h/2 A_k)^{-1}(x_l(t_n) + \begin{bmatrix} B_k^1 & \vdots & B_k^2 \end{bmatrix} \begin{bmatrix} u_l(t_{n+1/2}) \\ \cdots \\ y_{nl}(t_n) \end{bmatrix}). \quad (7)$$

Stage 2: The system integrates from midpoint $t_{n+1/2}$ to next point $t_{n+1}=t_n+h$.

Stage 2-1: NL and PWL parts parallelly calculate the interface variables $y_l$ and $y_{nl}$ at time point $t_{n+1/2}$ through (2) and (5).

Stage 2-2: Upon obtaining the interface variables, the NL and PWL parts are stepped in parallel from $t_{n+1/2}$ to $t_n$ through

$$x_{nl}(t_{n+1}) = x_{nl}(t_n) + h f_{nl}(x_{nl}(t_{n+1/2}), y_l(t_{n+1/2})), \quad (8)$$

$$x_l(t_{n+1}) = x_l(t_n) + h(A_k x_l(t_{n+1/2}) + \begin{bmatrix} B_k^1 & \vdots & B_k^2 \end{bmatrix} \begin{bmatrix} u_l(t_{n+1/2}) \\ \cdots \\ y_{nl}(t_{n+1/2}) \end{bmatrix}). \quad (9)$$

In the Stage 1, the explicit method as (6) are used to solve the NL parts to avoid solving NL equations iteratively, and the implicit method as (7) are used to solve the PWL parts to increase the overall numerical stability. In should be noted that the first stage is similar to the conventional latency-based methods since the NL parts use the interface variables at current time point $t_n$ for integration.

In the Stage 2, taking advantage of the derived values at midpoint $t_{n+1/2}$, the NL parts and PWL parts are all integrated using the same explicit mid-point methods to eliminate the latency between the NNL and PWL parts. The first and second stages together constitute a complete integration step. The subsequent section will substantiate that the introduced stage guarantees a second-order numerical accuracy and enhanced numerical stability.

*C. Numerical Stability and Accuracy*

The stability regions of different methods are plotted in Fig. 4. Stability region of the algorithm can be analyzed using the following test equation

$$\dot{x} = \lambda_0 x + \lambda_1 x, \quad (10)$$

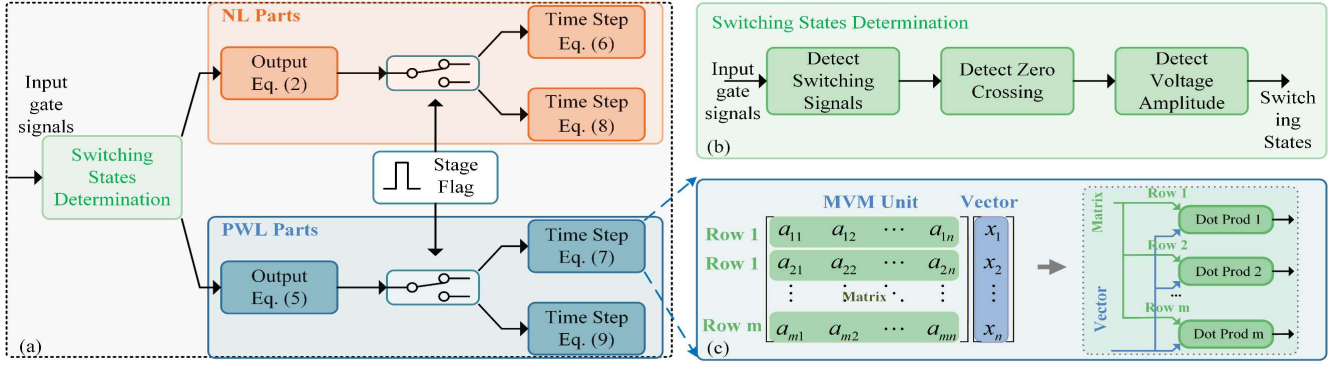

Fig. 5. IMEX Solver design. (a) Data-view of the IMEX solver. (b) Switching states determination process. (c) MVM Unit.

where $\lambda_0$ and $\lambda_1$ respectively correspond to the part computed using explicit and implicit methods. Applying the proposed solver to this test function can get

$$\frac{y_{n+1}}{y_n} = \frac{(z_0+1)^2 + 1 + z_1(z_0+1)}{2 - z_1}, \quad (11)$$

where $z_0=h\lambda_0$, $z_1=h\lambda_1$. For plotting the stability regions, $z_0$ is fixed, and the stability regions of $z_1$ are plotted in Fig. 4 under different values of $z_0$, which satisfy A-stable criterion. In fact, this method is A-stable, as long as $z_0$ satisfies $|z_0+1|<1$. On the contrary, the stability regions of explicit methods are limited.

For analyzing the numerical accuracy, a complete integration step and a single-variable dynamic system are considered.

$$\dot{x} = f_{nl}(x) + f_l(x), \quad (12)$$

where the subscript nl and l denote a nonlinear and linear function. Applying the proposed solver to this system can get

$$x_{n+1} = x_n + h(f_{nl}(x_n) + f_l(x_n)) + \frac{h^2}{2}(f_{nl}^{(1)}(x_n) + f_l^{(1)}(x_n))\dot{x}_n + O(h^3) \quad (13)$$

Based on the above analysis, this solver has second-order numerical accuracy.

## IV. FPGA IMPLEMENTATION

### A. IMEX Solver Design

This section introduces the implementation of the proposed IMEX method on the FPGAs. As shown in Fig. 5(a), the switching states of switches in the PWL parts should be determined before doing numerical integration. The actual switching states depend on both the external switching signals given by physical controllers and the values of the variables in the FPGA-based simulator. The switching states determination process is shown in Fig. 5(b). After determining the switching states, the PWL and NL parts are solved parallelly and choose the right computational unit based on current stage flag.

To solve the PWL parts, the TAMP method in [21] is adopted. Since the single-step computation of PWL parts of this method involves only matrix vector multiplications (MVMs), the MVM unit is designed to fully exploit the fine-grained parallel computing characteristic of the FPGA, as shown in Fig. 5(c). In terms of the NL parts, the magnetic coupling components are influenced by the operation conditions (e.g., the receiver coils' positions and velocity). The operation conditions can either be predefined on FPGAs or sent to FPGAs via external signals. Based on the operation conditions, the corresponding inductance values are retrieved from the look-up tables in FPGAs.

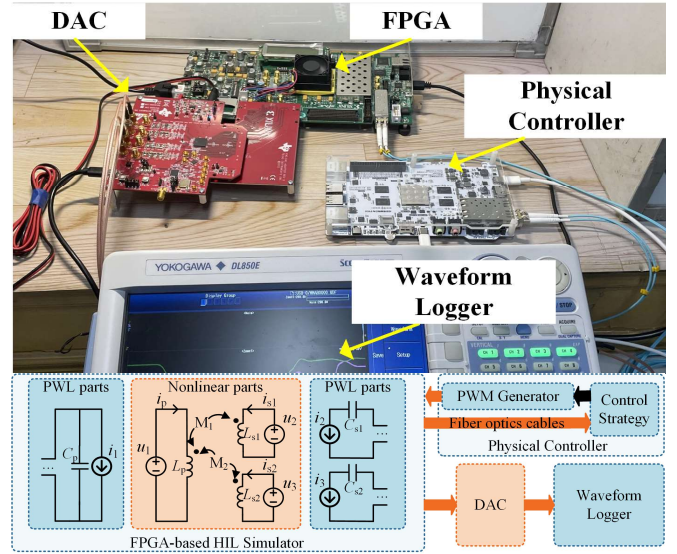

Fig. 6. RT-HIL hardware platform.

### B. Simulation Platform

The proposed solver is implemented on an FPGA computing platform using high-level synthesis to achieve RT-HIL simulation. As shown in Fig. 6, the FPGA computing platform is based on Xilinx's VC707 evaluation board, which houses the XC7VX485T-2FFG1761C chip. The XC7VX485T FPGA includes 303600 look-up tables (LUTs), and 2800 DSP slices. The FPGA real-time simulator and the physical controller employ an optical fiber communication module featuring an SFP/SFP+ module connector which enables the FPGA simulator to both read signals from the optical fiber module and transmit signals to it. The Ethernet communication rates can be up to 1000Mb/s. This platform also equips with a digital-to-analog converter to display the simulated waveforms on a waveform logger. TI's quad-channel 16-bit DAC34H84 evaluation board with data rates of up to 1.25 GSPS is selected for the DAC.

### C. FPGA Implementation Details

Fixed-point format is utilized in our FPGA implementation to further accelerate the computations. Compared to floating-point format, operations in the fixed-point format involve

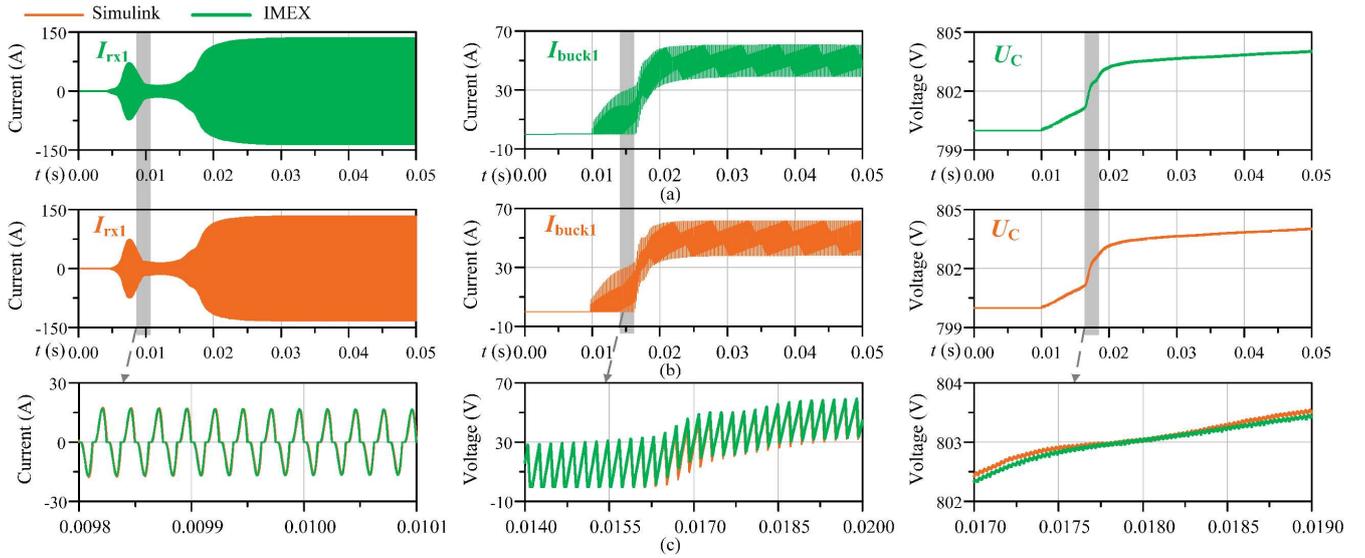

Fig. 7. Simulation results of the starting process of the WPT system. (a) The simulation results of the proposed IMEX method. (b) The simulation results of Simulink®. (c) Zoomed in view of the comparison between the simulation results of the proposed methods and Simulink®.

TABLE I
MAIN PARAMETERS OF WPT SYSTEM

| Parameters | Value |
|---|---|
| $U_{in}$ | 1.5kV |
| $L_{fl}$ | 42.95μH |
| $C_p$, $C_{s1}$, $C_{s2}$ | 442.8nF, 62.34nF, 62.34nF |
| $C_{f1}$, $C_{f2}$ | 400μF |
| $L_{B11}$, $L_{B12}$, $L_{B21}$, $L_{B22}$ | 1.1mH |
| $f_{sw}$ (transmitter, receiver) | 40kHz, 5kHz |

TABLE II
TYPICAL VALUES OF THE SIMULATION RESULTS

| Variables | Value Type | IMEX | Simulink | Relative Error |
|---|---|---|---|---|
| $I_{rx1}$ | RMS value /A | 93.44 | 94.93 | 1.57% |
|  | Peak value /A | 132.9 | 134.9 | 1.48% |
| $U_{tx}$ | RMS value /V | 2658 | 2690 | 1.19% |
|  | Peak value /V | 3668 | 3709 | 1.11% |
| $U_C$ | RMS value /V | 959.9 | 942.2 | 1.88% |
|  | Peak value /V | 960.6 | 943.0 | 1.87% |
| $I_{buck11}$ | RMS value /A | 49.32 | 49.96 | 1.28% |
|  | Peak value /A | 61.39 | 61.80 | 0.66% |
| $I_{out}$ | RMS value /A | 200.0 | 199.9 | 0.05% |
|  | Peak value /A | 201.1 | 201.0 | 0.05% |

*Measurement time range of steady state value: 0.048s~0.049s
**1) $I_{rx1}$ represents the receiving coil current; 2) $U_{ctx}$ represents the bus capacitor voltage at the transmitting side; 3) $U_C$ represents the bus capacitor voltage of the buck circuit; 4) $I_{buck11}$ represents Buck1 output current; 5) $I_{out}$ represents the supercapacitor charging current. (Also labeled in Fig. 1.)

simpler arithmetic calculations, primarily focused on addition and bit manipulation. Nonetheless, the fixed-point format exhibits limited numerical precision, which can be overcome by meticulously configuring the number of integer and fractional bits.

In our case, a fixed-point format featuring a width of 64 bits and an integer bit-width of 24 bits is chosen. This choice guarantees a level of numerical accuracy comparable to that achieved through floating-point operations. To program the FPGA, the C++ source code of the TA-MP method is tailored to meet the precise prerequisites of the study, subsequently translated into hardware description language (HDL) through employment of the Vitis® HLS® tool.

Subsequently, the HDL code undergoes optimization for both timing and resource utilization, employing high-level synthesis (HLS) instructions. The culminating step involves loading the comprehensive design, encompassing other functional modules, onto the FPGA hardware utilizing Xilinx®'s Vivado® tool.

IV. CASE STUDY

This section applies the IMEX solver to simulate the studied railway WPT system shown in Fig. 1. The system's parameters are shown in Table I. The accuracy of the proposed method is verified by comparing with commercial simulation software using pure an implicit method. Moreover, the method is compared with the conventional latency-based methods to verify its superior numerical stability. Then, the proposed solver is implemented on an FPGA to realize the RT-HIL simulation. Lastly, a comprehensive evaluation on the proposed method is conducted.

A. Accuracy Verification

To verify the simulation accuracy of the IMEX method, the simulation results are compared with the results of using Simulink®. The IMEX method is implemented using C++ code in the S-function model and adopts a step-size of 75ns. This step-size is the same as the step-size used in the following FPGA implementation. We also used the existing components in the Simscape library to build the WPT system. For simulating this WPT system, the integration method used in Simulink is ode23s, an implicit variable-step method that can adaptively change step-sizes to meet accuracy requirements.

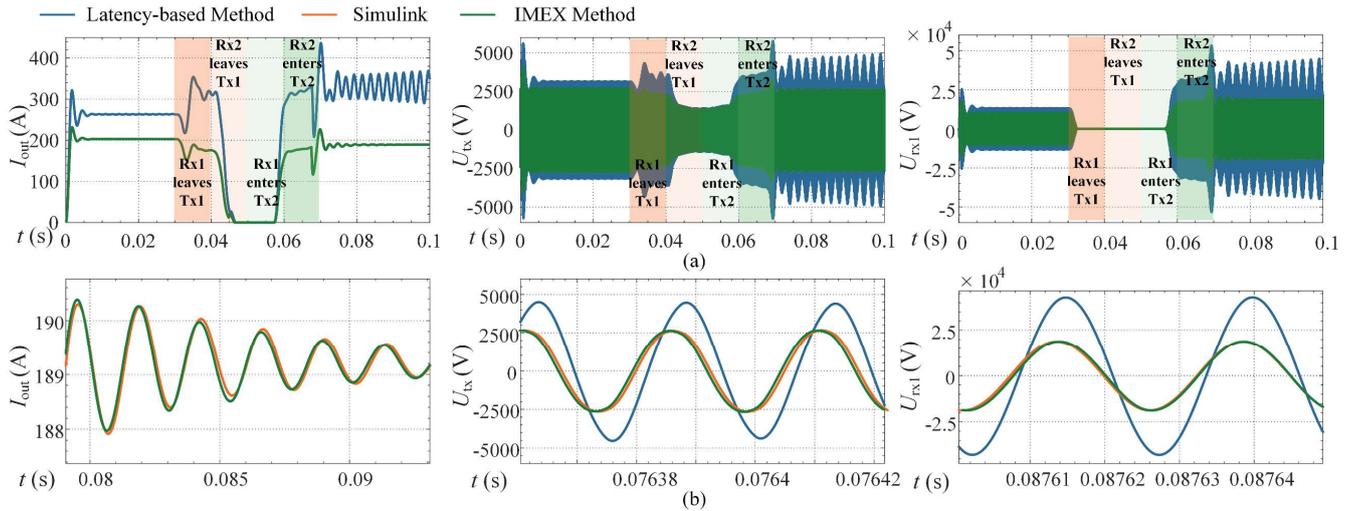

Fig. 8. Simulation results of the dynamic process of the WPT system under open-loop control. (a) The simulation results of the proposed IMEX method, latency-based method, and Simulink®. (b) Zoomed in view of the simulation results comparison.

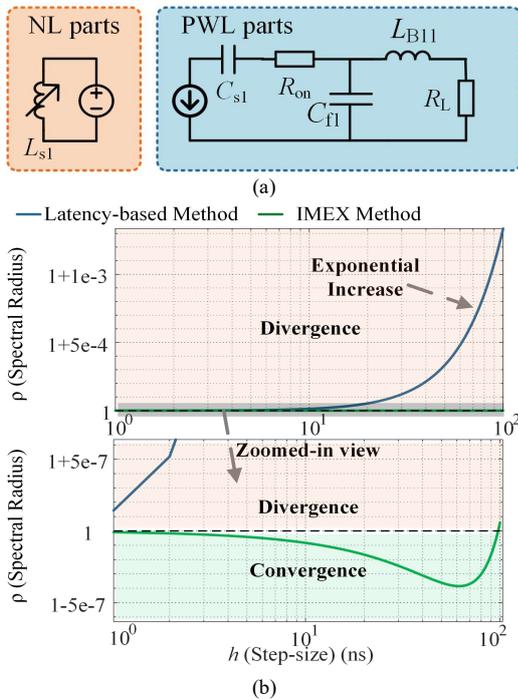

Fig. 9. Stability comparison of the IMEX method and conventional latency-based method. (a) Simplified study circuit. (b) Comparison of spectral radius of different methods' discretized matrices.

The starting process of the WPT system is simulated. Under this condition, the receiver coils are assumed to be stationary, and inductance values are constant. The transmitter starts by phase shifting within 0.01s and shift phase is not adjusted after starting condition. The receiver does not work when the transmitter is phase-shifted, and starts to work after 0.01s, using closed-loop strategy to control the average value of the output current.

Fig. 7(a) shows the simulation results of Simulink®, Fig. 7(b) shows the simulation results of the IMEX method, and Fig. 7(c) shows the enlarged view of the comparison between the simulation results of Simulink® and the IMEX method. Among the waveforms, $I_{rx1}$ represents the receiving inductor current, $I_{buck1}$ represents the buck output current, and $U_c$ represents the supercapacitor charging voltage. These variables are also highlighted in Fig. 1(d).

Typical values of the simulation results are compared in Table II. Relative error of steady-state results below 2%. It shows that the simulation results of the IMEX method and Simulink® are very close, even in dynamic processes.

### B. Compared with Latency-based Method

In this section, we undertake a comparative analysis between the IMEX method and the traditional latency-based method in order to show the better numerical stability of the former. As mentioned in Section III-B, the first stage of the IMEX method is identical to the conventional latency-based method. Consequently, implementing the conventional latency-based method is simply only remaining the first stage of the proposed IMEX method.

The system is simulated under the conditions where the train maintains a consistent velocity, and the two receiver coils, one in front and one in back, transit from one transmitter coil to the subsequent one. As a result, the inductance values undergo dynamic changes. Within this context, an open-loop control strategy is employed to preclude any influence stemming from the control component on the simulation outcomes, so that the stability of these methods' solving circuits can be straightforwardly compared.

The simulation results are shown in Fig. 8. The waveforms given by the IMEX method agree with the reference waveforms given by Simulink®. However, the waveforms generated by the traditional latency-based method manifest substantial disparities. This phenomenon can be ascribed to the inherent numerical instability of the latency-based method.

In order to demonstrate the superior numerical stability of the IMEX method more clearly, we scrutinize a simplified circuit illustrated in Fig. 9(a). This simplified circuit is part of the WPT system corresponding to a specific switching state. After applying the IMEX and latency-based methods to solve this circuit, two discretized matrices are derived. These discretized matrices depend on the step-sizes. Should the spectral radius of a discretized matrix surpass 1, the simulation outcomes will diverge and exhibit unbounded growth. When it is less than 1, the simulation results converge, and the corresponding method is stable under this condition.

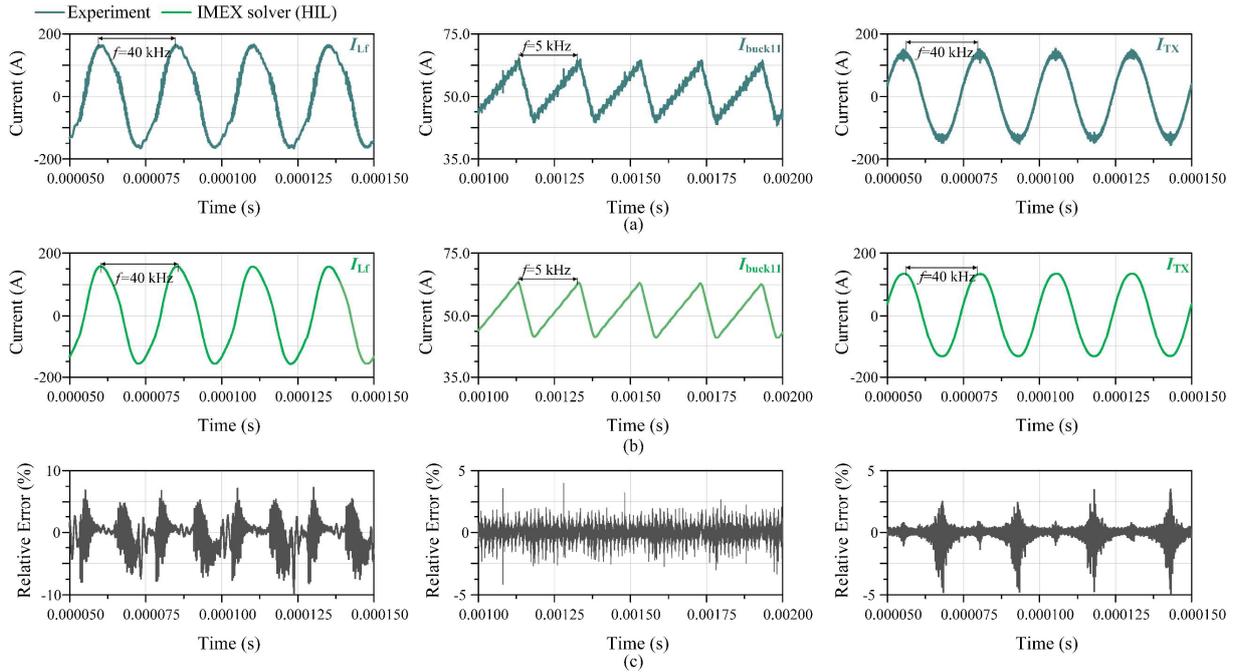

Fig. 10. Comparison between the IMEX solver HIL waveform and experiment waveform. (a). Experiment waveform. (b) IMEX solver HIL waveform (c). Relative errors.

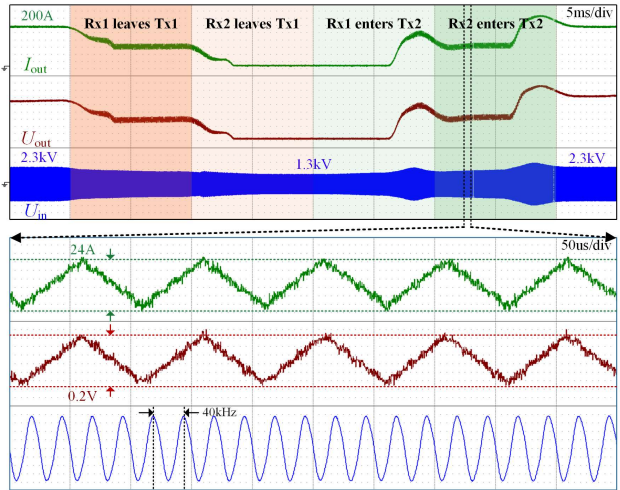

Fig. 11. RT-HIL simulation results.

TABLE III
FPGA RESOURCE UTILIZATION AND RELATIVE ERROR

| Methods | LUTs | DSP48s | Calculation Time | Relative Error |
|---|---|---|---|---|
| IMEX | 21119 (6.96%) | 439 (15.68%) | 68.212ns (<75ns) | 0.021% |
| Trapezoidal Method | 24231 (7.98%) | >100% | 22418.877ns (>>75ns) | 0.022% |
| Forward Euler Method | 19212 (6.33%) | 305 (10.90%) | 64.527ns (<75ns) | Diverge |
| Latency-based Method | 19534 (6.43%) | 320 (11.45%) | 64.229ns (<75ns) | Diverge |

The spectral radius of two discretized matrices comparison is shown in Fig. 9(b). As discernible, the spectral radius of the discretized matrix associated with the latency-based method consistently exceeds 1 across varying step sizes, elucidating the tendency towards infinite results as observed in Fig. 8. On the contrary, the spectral radius corresponding to the IMEX method is less than 1 when the step-size is smaller than 100ns as shown in the zoomed-in view in Fig. 9(b).

*C. RT-HIL Simulation Results*

In this section, we implement the IMEX algorithm on an FPGA to enable RT-HIL Simulation. Firstly, the static conditions where the receiver coils are stationary are simulated. The simulated results are compared with the experimental results of a prototype for validation. Fig. 10 shows the currents on the transmitter side ($I_{Lf}$, $I_{TX}$), the current on the receiver side ($I_{buck11}$), and the relative error between the results.

Then the dynamic conditions where the receiver coils are in movement are simulated. Fig. 11 shows the results of the RT-HIL simulation, including the waveforms of the output current ($I_{out}$) and voltage ($U_{out}$), and the input voltage ($U_p$) when the receiver coils Rx1 and Rx2 switch between different transmitter coils. The current and voltage drop when the receiver coil leaves a transmitting coil due to the lack of energy transfer. Then, after entering the next transmitting coil, the output current and voltage stabilize again due to closed-loop control of the controller.

The FPGA resource utilization results and comparisons are shown in Table III. The proposed solver consumes 21119 LUTs and 439 DSPs. The consumed computational resource (e.g., DSP48s) of the latency-based method and forward Euler method (an explicit method) is less than that of the IMEX method. Because, the computational burden of latency-based method and forward Euler method are close to one stage of the IMEX method. However, the latency-based method and forward Euler method are both unstable when simulating the studied WPT system due to their lack of stability. It's worth noting that even though the IMEX method has one more stage, the computational delay is halved by its implementation design where each computational step only calculates one stage, and two stages are calculated alternatively. Trapezoidal method, as a pure implicit method, is stable but takes up too many resources to be implemented on the FPGA board since iteratively solving NL equations is required.

*D. Comprehensive Evaluation*

Through the above analysis and experiments, it can be concluded that the proposed IMEX solver successfully addresses the challenges of simulating the WPT system.

Firstly, this solver has large numerical stability regions in the implicit part, which adapts well to the stiffness of the PWL parts.

In addition, this solver solves NL parts explicitly, eliminating the need for iterative solutions of nonlinear equations, so that small step-sizes can be achieved. Even though conventional latency-based methods also avoid solving NL equations to achieve small step-sizes, the proposed method has superior stability, which is achieved by an additional integration stage that eliminates the latency between NL and PWL parts.

Lastly, dividing one integration step into two stages allows each computational step to only calculate one stage, so that computational delays are further reduced. Furthermore, the parallel computation of the NL and PWL parts takes advantage of the parallel computing capabilities of FPGAs, further reducing delays.

As a result, the proposed simulator successfully simulated the studied WPT system with a step-size of 75ns. The small step-size of this solver, together with the fast communication speed between this simulator and controller, collectively enable comprehensive testing of controllers for the WPT system such as controller communication testing.

## V. CONCLUSION

An FPGA-based IMEX simulation solver for solving railway WPT systems with NL magnetic coupling components is proposed. The solver partitions the system into PWL and NL parts and solves them with different methods parallelly. This solver avoids solving NL equations by solving NL parts explicitly and achieves high numerical stability by solving PWL parts implicitly. Compared with pure implicit methods, this solver saves up computational resources, so that larger-scale systems can be simulated on the same platform. The added stage in this solver eliminates the latency and increases the stability, differing it from the conventional latency-based methods. As a result, this solver achieves step-size of 75ns on an FPGA-based HIL simulator to accurately solve a WPT system with 28 switches. Moreover, this method has the potential to be applied to solve other power electronic systems that contain NL components.